\documentstyle[prl,aps,epsf,bm,twocolumn]{revtex}

\ifpreprintsty
  \newcommand{\pathint}{{ \int \hspace{-0.95em} \rule[-1.45ex]{0.65em}{0.1ex} \hspace*{0.0em}\rule[2.5ex]{0.65em}{0.1ex} \hspace*{-0.95em} \int \hspace{-0.25em}}}
\else
  \newcommand{\pathint}{{ \int \hspace{-1.1em} \rule[-1.8ex]{0.65em}{0.1ex} \hspace*{0.15em}\rule[2.9ex]{0.65em}{0.1ex} \hspace*{-1.05em} \int \hspace{-0.25em}}}
\fi

\begin{document}
\draft
\ifpreprintsty \else \wideabs{ \fi
\title{On the approximation of Feynman-Kac path integrals for quantum statistical mechanics}
\author{Stephen D. Bond}
\address{Department of Mathematics, University of Kansas, 
Lawrence, KS 66045, USA}
\author{Brian B. Laird }
\address{ Department of Chemistry, University of Kansas, 
Lawrence, KS 66045, USA}
\author{Benedict J. Leimkuhler}
\address{Department of Mathematics \& Computer Science, 
University of Leicester, Leicester LE1 7RH, UK}
\maketitle
\begin{abstract}
Discretizations of the Feynman-Kac path integral representation of the quantum 
mechanical density matrix are investigated.  Each infinite-dimensional path 
integral is approximated by a Riemann integral over a finite-dimensional 
function space, by restricting the integration to a subspace of all admissible
 paths.  Using this process, a wide class of methods can be derived, with each 
method corresponding to a different choice for the approximating subspace.  
The traditional ``short-time'' approximation and ``Fourier discretization'' 
can be recovered from this approach, using linear and spectral basis 
functions respectively.  As an illustration, a novel method is formulated 
using cubic elements and is shown to have improved convergence 
properties when applied to a simple model problem.
\end{abstract} 
\ifpreprintsty \else } \fi

The path integral approach provides a powerful method for studying properties 
of quantum many-body systems \cite{FeHi:65}.  When applied to statistical 
mechanics \cite{Fe:72}, each element of the quantum density matrix is 
expressed as an integral over all curves connecting two configurations:
\begin{equation}
\rho\left({\mathbf b} , {\mathbf a} \right) = 
\pathint_{{\mathbf a}:{\mathbf b}} 
{\cal D}\left[{\mathbf x}\left(\tau\right)\right] 
\exp \left\{ - \frac{1}{\hbar} \Phi\left[ {\mathbf x}\left(\tau \right) ; 
\beta \right] \right\}. \label{dense_mat}
\end{equation}
The symbol ${\cal D}\left[ {\mathbf x} \left( \tau \right) \right]$ indicates 
that the integration is performed over the set of all differentiable curves, 
${\mathbf x}:[0,\beta \hbar] \rightarrow {\mathbf R}^{d}$, with
${\mathbf x}\left(0\right) = {\mathbf a}$ and 
${\mathbf x}\left( \beta \hbar \right) = {\mathbf b}$.  The integer $d$ 
reflects the dimensionality, with $d = 3N$ for a system of $N$-particles in 
3-dimensional space.  The functional $\Phi$ can be derived from the classical 
action by introducing a relationship between temperature and imaginary time 
($ i t =  \beta \hbar $) \cite{FeHi:65}.  In this Letter, we will restrict 
our attention to the quantum many-body system, for which $\Phi$ takes the 
following form:
\begin{equation}
\Phi\left[ {\mathbf x}\left(\tau\right) ; \beta \right] = \int_0^{\beta \hbar}
 \frac{1}{2} \sum_{i=1}^{d} m_i \, \dot{x}_i\left(\tau\right)^2 +  
V\left[ {\mathbf x}\left(\tau\right)\right] {\mathrm  d}\tau.
\label{action-functional}
\end{equation}

Calculating the path integral in (\ref{dense_mat}) is a challenging task, 
which in general cannot be performed analytically.  It is only for simple 
model problems, such as quadratic potentials that an exact solution can be 
obtained.  For more challenging systems, the path integral has traditionally
been estimated using either the ``short-time'' approximation (STA) 
\cite{ScStChWo:81} or ``Fourier discretization'' (FD) \cite{Mi:75,FrDo:84}.  
Many authors have proposed improvements to the standard STA and FD, using 
techniques such as improved estimators \cite{HeBrBe:82}, 
partial averaging \cite{DoCoFr:85,CoFrDo:86,CoFrDo:89}, higher-order 
exponential splittings \cite{DRDR:83}, advanced reference 
potentials \cite{TaIm:84}, semi-classical expansions 
\cite{MaMi:89}, and extrapolation \cite{MiSrTr:00}.  The fundamental approach 
is the same in all of these methods:  the path integral is reduced to a high 
(but finite) dimensional Riemann integral, which is approximated using either 
a Monte Carlo or Molecular Dynamics.

The aim of this Letter is to provide a framework for the formulation of a wide
 class of methods for the discretization of quantum mechanical path integrals.
The idea of approximating path integrals using a finite subset of basis 
functions has been suggested before in the literature.   Davison was one of 
the first to consider the use of orthogonal function expansions in the 
representation of Feynman path integrals \cite{Da:54}, although he did not
explore truncating the expansion.  In a related article on 
Wiener integration, Cameron proposed using a finite set of orthogonal basis 
functions, and investigated the convergence of Fourier (spectral) elements 
\cite{Ca:51}.    In this Letter, we do not require that the basis functions 
are orthogonal, allowing for the direct comparison of the STA and FD methods.  
Although other authors have explored fundamental connections between the STA 
and FD \cite{Co:86} methods, we are unaware of any comparison using the  
approach investigated here.  In addition, our approach allows for the 
construction of new methods using general classes of orthogonal polynomials 
or finite elements.  To illustrate the flexibility of this approach, we 
derive a new method, using compactly supported (Hermite) cubic splines (HCS), 
which is shown to exhibit improved efficiency when applied to 
model problems.

To illustrate how one can use a subspace approximation to discretize the 
quantum density matrix in (\ref{dense_mat}), we start by introducing a change 
of variables to simplify the boundary conditions and temperature dependence 
for each path integral: ${\mathbf x}\left(\tau \right)  =  {\mathbf a} + 
\left({\mathbf b} - {\mathbf a}\right) \tau / \beta \hbar  + 
{\mathbf y}\left( \tau / \beta \hbar \right)$. 
Since the admissible paths, ${\mathbf x}$, satisfy the boundary conditions 
${\mathbf x}\left(0\right) = {\mathbf a}$ and 
${\mathbf x}\left(\beta \hbar \right) = {\mathbf b}$, the reduced paths given 
by ${\mathbf y}$, will satisfy Dirichlet boundary conditions, 
${\mathbf y}\left(0\right) = {\mathbf y}\left( 1 \right) = {\mathbf 0}$, 
independent of ${\mathbf a}$, ${\mathbf b}$, and $\beta$.  Introducing this 
change of variables into (\ref{dense_mat}), results in the following:
\ifpreprintsty
\begin{equation}
\rho\left({\mathbf b} , {\mathbf a} \right)   =  \pathint_{0:0} 
{\cal D}\left[{\mathbf y}\left(\frac{\tau}{\beta \hbar} \right)\right] 
 \exp \left\{ - \frac{1}{\hbar} \Phi\left[ {\mathbf a}  
 + \left({\mathbf b} - {\mathbf a} \right) \frac{\tau}{\beta \hbar} + {\mathbf y}\left( \frac{\tau}{\beta \hbar}  \right) ; \beta \right] \right\}. 
\label{density_matrix}
\end{equation}
\else
\begin{eqnarray}
\lefteqn{ \rho\left({\mathbf b} , {\mathbf a} \right)   =  \pathint_{0:0} 
{\cal D}\left[{\mathbf y}\left(\frac{\tau}{\beta \hbar} \right)\right] \times } \nonumber \\*
& & \exp \left\{ - \frac{1}{\hbar} \Phi\left[ {\mathbf a}  
 + \left({\mathbf b} - {\mathbf a} \right) \frac{\tau}{\beta \hbar} + {\mathbf y}\left( \frac{\tau}{\beta \hbar}  \right) ; \beta \right] \right\}. 
\label{density_matrix}
\end{eqnarray}
\fi
Note that the $i$th component of each reduced path ${\mathbf y}$, denoted by 
$y_i$, is a real-valued function on the interval $[0,1]$, satisfying Dirichlet
 boundary conditions.  For the systems considered in this article, we also 
require that the derivative of each $y_i$ is measurable 
(i.e., square-integrable).  Functions of this form are 
members of an infinite dimensional Sobolev space \cite{StFi:73}, defined by
$ {\cal S}_0^1\left[0,1\right] = \left\{ w \in {\cal C}\left[0,1 \right] \left| w\left(0\right) = w\left(1\right) = 0  \; {\mathrm  and } \; \| w \|_{\cal S} < \infty \right. \right\} , $ where  $\| w \|^2_{\cal S} \equiv \int_0^{1} \dot{w}\left(\xi \right)^2 + w \left( \xi \right)^2 {\mathrm d}\xi. $

We proceed in the following manner to discretize (\ref{density_matrix}):  
Consider a sequence of subspaces of increasing 
dimension ${\cal V}_1, \cdots, {\cal V}_{P}, \cdots \subset 
{\cal S}^{1}_0\left[0,1\right]$, where each ${\cal V}_P$ is of dimension 
$P$.  For convenience, let each subspace be defined as the span of a 
particular set of basis functions: ${\cal V}_P = {\mathrm  span} 
\left\{ \psi_1, \cdots, \psi_{P} \right\}$.  Now, given a component 
function $y_i \in {\cal S}^{1}_0\left[0,1 \right]$, we can define its 
projection on ${\cal V}_{P}$ uniquely by
\[
y^{(P)}_{i}\left(\xi\right) \equiv \sum_{k=1}^{P} \alpha_{k,i} 
\psi_k\left(\xi \right) \; , 
\]
Using the projection $y_i^{(P)}\left(\xi\right)$ as an approximation of 
$y_i\left(\xi\right)$ reduces the infinite-dimensional path integral in 
(\ref{density_matrix}) to a finite-dimensional Riemann integral over the 
coefficients, $\alpha_{i,k}$:
\ifpreprintsty
\begin{equation}
\tilde{\rho}\left({\mathbf b} , {\mathbf a} \right) 
 =  \int {\mathrm d}\bm{\alpha} \, J \, \exp \left\{ - \frac{1}{\hbar} 
\Phi\left[ {\mathbf a} + \left({\mathbf b} - {\mathbf a} \right) 
\frac{\tau}{\beta \hbar} 
+ {\mathbf y}^{(P)}\left(\frac{\tau}{\beta \hbar} \right) ; 
\beta \right] \right\}. \label{approx_density}
\end{equation}
\else
\begin{eqnarray}
\tilde{\rho}\left({\mathbf b} , {\mathbf a} \right) 
& = & \int {\mathrm d}\bm{\alpha} \, J \, \exp \left\{ - \frac{1}{\hbar} 
\Phi\left[ {\mathbf a} + \left({\mathbf b} - {\mathbf a} \right) 
\frac{\tau}{\beta \hbar} \right. \right. \nonumber \\*
& & \left. \left. + {\mathbf y}^{(P)}\left(\frac{\tau}{\beta \hbar} \right) ; 
\beta \right] \right\}. \label{approx_density}
\end{eqnarray}
\fi
Here, we have used simplified notation for the multi-dimensional integral, 
with ${\mathrm  d} \bm{ \alpha } \equiv \prod_{k,i} {\mathrm  d}\alpha_{k,i}$.
  The constant $J$ reflects the particular choice of variables, and can be 
readily calculated (as discussed later).  The reader should note 
that (\ref{approx_density}) does not depend on the basis functions chosen to 
represent the approximating subspace.  If both $\{\psi_1,\cdots,\psi_{P}\}$ 
and $\{\tilde{\psi}_1,\cdots,\tilde{\psi}_{P}\}$ span ${\cal V}_P$, then 
there is an invertible linear transformation (i.e., change of variables) 
${\mathbf U}$ such that 
$\tilde{\bm{ \alpha } } = {\mathbf U} \bm{ \alpha } $.

To show in detail how subspace methods can be applied in practice, we 
consider the case of an $N$-body Hamiltonian system:
\[
\hat{H} = \frac{1}{2} \sum_{i=1}^{d} m_i \, \hat{p}_i^2 + 
V\left[x_1, \cdots, x_d \right] \; . 
\]
Here, the coordinate and momentum operators are denoted by $x_i$ and 
$\hat{p}_i$ respectively.  For this system the functional $\Phi$ is given 
by (\ref{action-functional}), which when applied to the projected path, 
${\mathbf x}^{(P)}\left(\tau\right) \equiv {\mathbf a} + ({\mathbf b} - 
{\mathbf a}) \tau/\beta \hbar + {\mathbf y}^{(P)}
\left(\tau/\beta \hbar \right)$, results in
\[
\Phi\left[{\mathbf x}^{(P)}\left(\tau\right) ; \beta \right] = 
\int_0^{\beta \hbar} \sum_{i=1}^{d} \frac{m_i}{2}  
\left[ \dot{x}^{(P)}_{i}\left(\tau\right) \right]^2 + 
V\left[ {\mathbf x}^{(P)}\left(\tau\right) \right] 
{\mathrm  d}\tau \; . 
\]
After expanding the $\tau$-integrals, introducing a change of variables 
$\xi = \tau/\beta \hbar $, and using the boundary conditions of each 
$\psi_k$, we have
\ifpreprintsty
\begin{eqnarray}
\Phi\left[{\mathbf x}^{(P)}\left(\beta \hbar \, \xi\right) ; \beta \right] & = & \sum_{i=1}^{d} \frac{m_i}{2 \, \beta \, \hbar} \left\{ \, 
\left( b_i - a_i \right)^2  
+ \vec{\alpha}^T_{i} {\mathbf K} 
\vec{\alpha}_{i} \right\} \nonumber \\* 
& & \hspace{1cm} + \beta \hbar \int_0^{1} 
V\left[ {\mathbf a} + ({\mathbf b} - {\mathbf a}) \xi + 
{\mathbf y}^{(P)}\left(\xi\right) \right] {\mathrm  d}\xi \; , 
\label{action-approx3}
\end{eqnarray}
\else
\begin{eqnarray}
& & \Phi\left[{\mathbf x}^{(P)}\left(\beta \hbar \, \xi\right) ; \beta \right] = 
\sum_{i=1}^{d} \frac{m_i}{2 \, \beta \, \hbar} \left\{ \, 
\left( b_i - a_i \right)^2  
+ \vec{\alpha}^T_{i} {\mathbf K} 
\vec{\alpha}_{i} \right\} \nonumber \\* 
\lefteqn{ \hspace{1cm} + \beta \hbar \int_0^{1} 
V\left[ {\mathbf a} + ({\mathbf b} - {\mathbf a}) \xi + 
{\mathbf y}^{(P)}\left(\xi\right) \right] {\mathrm  d}\xi \; , } 
\label{action-approx3}
\end{eqnarray}
\fi
where $\vec{\alpha}_i \equiv \left[ \alpha_{1,i} \cdots \alpha_{P,i} \right]^T$.
The ``stiffness matrix'', ${\mathbf K} \in {\mathbf R}^{P \times P }$, has 
entries given by the inner-product $K_{j,k} = \int_0^{1} 
\dot{\psi}_{j}\left(\xi\right) \dot{\psi}_{k}\left(\xi\right) 
{\mathrm  d}\xi $ .

Substituting (\ref{action-approx3}) into (\ref{approx_density}), we obtain a 
simplified expression for the approximate density matrix:
\ifpreprintsty
\begin{eqnarray}
\tilde{\rho}\left({\mathbf b} , {\mathbf a} \right) &=&  J \,
\exp \left\{ - \sum_{i=1}^{d}\frac{m_i}{2 \, \beta \, \hbar^2} 
\left( b_i - a_i \right)^2 \right\} \nonumber \\* 
&\times & \int {\mathrm  d}\bm{\alpha} \, 
\exp \left\{ - \sum_{i=1}^{d}\frac{m_i}{2 \, \beta \, \hbar^2} 
\vec{\alpha}^T_{i} {\mathbf K} \vec{\alpha}_{i} 
 - \beta \int_0^{1} V\left[ {\mathbf a} + ({\mathbf b} - {\mathbf a}) \xi + 
{\mathbf y}^{(P)}\left(\xi\right) \right] {\mathrm  d}\xi  \right\}, 
\label{approx_density2}
\end{eqnarray}
\else
\begin{eqnarray}
\tilde{\rho}\left({\mathbf b} , {\mathbf a} \right) &=&  
\exp \left\{ - \sum_{i=1}^{d}\frac{m_i}{2 \, \beta \, \hbar^2} 
\left( b_i - a_i \right)^2 \right\} \nonumber \\* 
& \times & \int {\mathrm  d}\bm{\alpha} \, J \, 
\exp \left\{ - \sum_{i=1}^{d}\frac{m_i}{2 \, \beta \, \hbar^2} 
\vec{\alpha}^T_{i} {\mathbf K} \vec{\alpha}_{i} \right. \nonumber \\*
& - & \left. \beta \int_0^{1} V\left[ {\mathbf a} + ({\mathbf b} - {\mathbf a}) \xi + 
{\mathbf y}^{(P)}\left(\xi\right) \right] {\mathrm  d}\xi  \right\}, 
\label{approx_density2}
\end{eqnarray}
\fi
For the Fourier case, one typically calculates $J$ by requiring that the discretization be exact when applied to an ideal gas (i.e., $V \equiv 0$) \cite{Mi:75,FrDo:84,Da:54}.  Applying this same technique to a generic subspace method, 
and assuming that ${\mathbf K}$ is positive definite, one can solve for $J$ 
in a straightforward manner:
\[
J =   \prod_{i=1}^{d} \sqrt{ \det {\mathbf K} } \, \left( \frac{m_i }{2 \, \pi \, \beta \, \hbar^2 } \right)^{\frac{P+1}{2}} \; .
\]

Before discussing particular choices for basis functions, we should mention 
that, in general, the one-dimensional $\xi$-integral in 
(\ref{approx_density2}) cannot be performed analytically.  This problem has 
been traditionally circumvented by using a discrete approximation, such as 
Gaussian quadrature \cite{FrDo:84,Co:86}.  For example, one can view the 
primitive STA as using the trapezoidal rule.   If the quadrature scheme is of
sufficiently high order its use will not reduce the asymptotic rate of 
convergence of the overall method.  An optimal scheme must be efficient, 
since for nonlinear $N$-body systems evaluating $V$ may be 
computationally expensive.

As mentioned above, the real benefit of using a general subspace approach 
is the flexibility afforded through the choice of basis functions.  
By considering a general class of pseudo-spectral or finite-element basis
functions, a diverse group of discretizations can be constructed.  Direct
comparisons can be made between basis functions of varying smoothness and 
support.  However, for brevity, we restrict our attention in this Letter 
to three different types of basis functions:  linear, spectral, and cubic 
elements.  Representative basis functions from each of these discretizations 
are shown in Figure \ref{elements}.

\ifpreprintsty
\else
\begin{figure*}[thp]
\centering
\epsfysize=2.40in 
\epsfbox{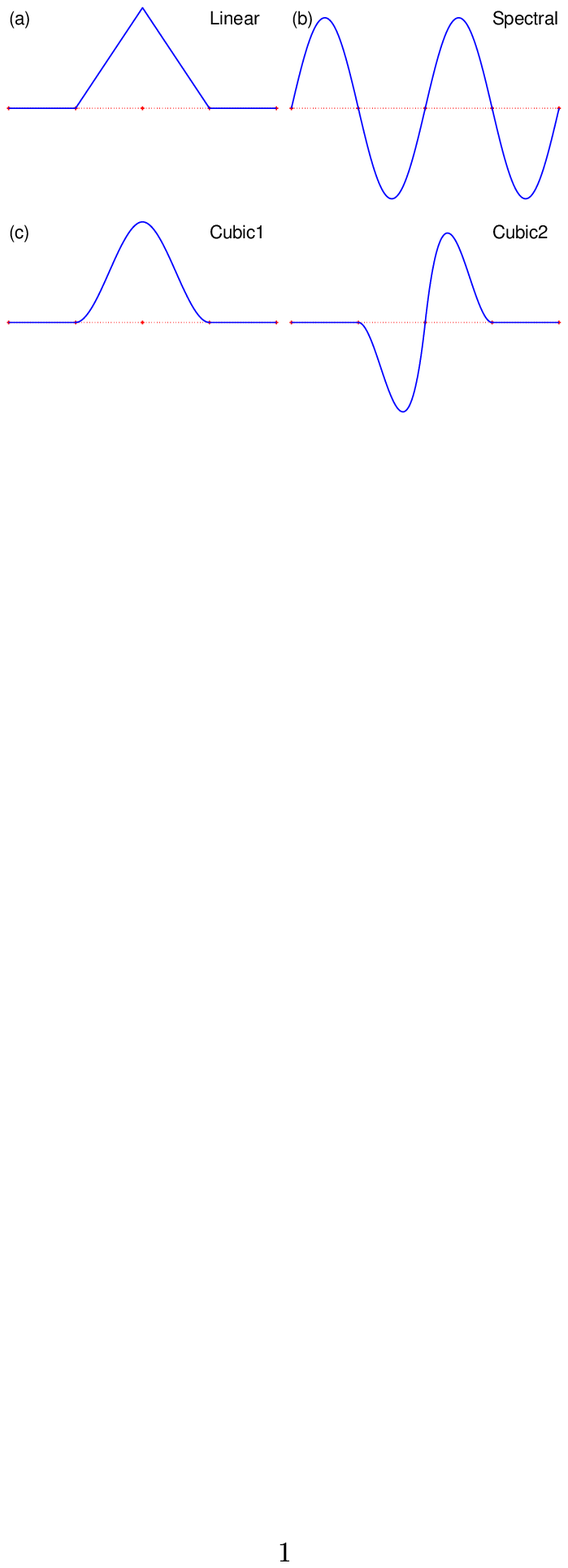}
\caption{Sample basis functions are shown above for the (a) linear, (b) spectral, and (c) cubic element methods.}
\label{elements}
\end{figure*}
\fi

The traditional STA method can be constructed by considering polygonal paths,
which can be represented by piecewise linear basis functions \cite{StFi:73}.  
For a given number of linear segments, $P+1$, we can define an approximating 
subspace ${\cal V}_{P}$ as the span of basis functions 
$\{ \psi_1, \cdots, \psi_{P} \}$, where each $\psi_k$ is defined by the 
following formula  
\begin{eqnarray*}
\psi_{k}\left(\xi\right) & := & \phi^{{\mathrm  lin}} \left( \xi \, (P+1) - k \right), 
\quad \\ {\mathrm  with} \quad
\phi^{{\mathrm  lin}} \left( u \right) & := & \left\{
\begin{array}{ccl}
1 - \left| u \right| & \quad  & u \in \left[ -1, 1 \right] \\*
0 & \quad & {\mathrm  otherwise} 
\end{array}
\right. .
\end{eqnarray*}
For this discretization, it is routine to show that the elements of the 
``stiffness'' matrix, ${\mathbf K} \in R^{P \times P}$, can be determined by
 $K_{i,j} = -1 \delta_{i-1,j} + 2 \delta_{i,j} - 1 \delta_{i+1,j}$.
In a similar manner, the FD method can be derived using the subspace approach 
by considering spectral basis functions of the form  
$\psi_k\left(\xi\right) = 1/k \sin \left(k \, \pi \xi \right)$.  ${\mathbf K}$ is diagonal for this basis, with entries given by 
$K_{i,j} = \pi^2/2 \, \delta_{i,j}$.

A new method can be constructed by approximating the space of paths using 
piecewise (Hermite) cubic splines (HCS) \cite{StFi:73}.  Each spline is 
defined on an interval of width $2/P$, with its shape uniquely determined
by its function value and derivative at the ends of the interval.  It is
assumed here that $P$ is an even integer.  Each piecewise cubic
path has a continuous derivative, and is described by linear combinations of 
the basis functions
\[
\psi_k = \left\{ \begin{array}{ccl} 
\phi^{{\mathrm  hcs}}_1\left(\xi P / 2 - k\right) & \quad & 1 \leq k < P / 2 \\
\phi^{{\mathrm  hcs}}_2\left(\xi P / 2 + P / 2 - k\right) & \quad & P / 2 \leq k \leq P  \end{array} \right. ,
\]
where
\begin{eqnarray*}
\phi^{{\mathrm  hcs}}_1\left(u\right) & := &
\left\{
\begin{array}{ccl}
\left(1 - \left| u \right| \right)^2 \left(2 \left| u \right| + 1\right) & \quad  & u \in \left[ -1, 1 \right] \\*
0 & \quad & {\mathrm  otherwise} 
\end{array}
\right.  
\quad \\
{\mathrm  and} \quad 
\phi^{{\mathrm  hcs}}_2\left(u\right) & := &
\left\{
\begin{array}{ccl}
u \left( 1 - \left| u \right| \right)^2 & \quad  & u \in \left[ -1, 1 \right] \\*
0 & \quad & {\mathrm  otherwise} 
\end{array}
\right.
.
\end{eqnarray*}
One can verify that the reduced path $y^{(P)}(\xi) = \sum \alpha_k \psi_k ( \xi ) $ satisfies
Dirichlet boundary conditions, and interpolates the interior grid points $(2 j / P, \alpha_j)$ for integers $1 \leq j < P/2$.  The derivative of the path at
all the grid points is determined by the remaining $P/2 + 1$ coefficients, $\alpha_k$.   Due to the compact support of the basis functions, the stiffness 
matrix is banded, with block structure:
\[
{\mathbf K}^{{\mathrm  hcs}} = \frac{P}{60}\left[ 
\begin{array}{ccc}
{\mathbf K}_1 & & {\mathbf K}_3\\
\\
{\mathbf K}^T_3 & & {\mathbf K}_2\\
\end{array}
\right] \; ,
\]
where the blocks are given by
\ifpreprintsty
\[
{\mathbf K}_1  =  \left[ 
\begin{array}{cccc}
72 & -36 & 0 & . \\
-36 & 72 & . & 0 \\
0 & . & 72 & -36 \\
. & 0 & -36 & 72 
\end{array}
\right] \; ,  \;
{\mathbf K}_2  = \left[ 
\begin{array}{cccc}
4 & -1 & 0 & . \\
-1 & 8 & . & 0 \\
0 & . & 8 & -1 \\
. & 0 & -1 & 4 
\end{array}
\right] \; ,
\, {\mathrm  and} \quad {\mathbf K}_3  =  \left[ 
\begin{array}{cccccc}
-3 & 0 & 3 & 0 & . & . \\
 0 & -3 & 0 & . & . & 0 \\
 0 & . & . & 0 & 3 & 0 \\
  . & . & 0 & -3 & 0 & 3 
\end{array}
\right] \; .
\]
\else
\begin{eqnarray*}
{\mathbf K}_1 & = & \left[ 
\begin{array}{cccc}
72 & -36 & 0 & . \\
-36 & 72 & . & 0 \\
0 & . & 72 & -36 \\
. & 0 & -36 & 72 
\end{array}
\right] \; , \quad \\
{\mathbf K}_2 & = &\left[ 
\begin{array}{cccc}
4 & -1 & 0 & . \\
-1 & 8 & . & 0 \\
0 & . & 8 & -1 \\
. & 0 & -1 & 4 
\end{array}
\right] \; , \\
\, {\mathrm  and} \quad {\mathbf K}_3 & = & \left[ 
\begin{array}{rrrrrr}
-3 & 0 & 3 & 0 & . & . \\
 0 & -3 & 0 & . & . & 0 \\
 0 & . & . & 0 & 3 & 0 \\
  . & . & 0 & -3 & 0 & 3 
\end{array}
\right] \; .
\end{eqnarray*}
\fi
Note that the blocks are not all the same size, with ${\mathbf K}_3$ of 
dimension $(P/2 - 1)\times (P/2 + 1)$.  The determinant of 
${\mathbf K}^{{\mathrm  hcs}}$ may be calculated exactly, but for most purposes 
it is enough to know that it is a constant, which will cancel out 
when (\ref{approx_density2}) is used to calculate averages.

As a numerical experiment, we apply each path integral discretization to the
problem of calculating the average energy of a particle in a one-dimensional
double-well.  We have chosen the same double-well potential considered in 
\cite{FrDo:84}, which is as 
follows: $V\left(x\right) = m \, \omega^2 \, x^2 / 2 + A/((x/a)^2 + 1)$.  The 
parameter values are all in atomic units, with $\omega = 0.006$, $A = 0.009$, 
$a = 0.09$, and $m = 1836$.  At low temperatures, the energy is just above 
$0.006$, which is below the barrier height of $0.009$.

To measure the accuracy of each method, we compute the energy at a fixed
temperature of $T = 0.1 \hbar \omega / k$, using Metropolis Monte Carlo to 
generate the canonically distributed configurations.  The one-dimensional 
line-integrals of the potential are approximated using Simpson's rule for the 
FD and HCS methods, and the traditional trapezoidal rule
for the STA method.  The number of integration nodes is set equal to the
number of basis functions, $P$, resulting in the same number of potential
evaluations for each method.  For the STA method this results the potential 
is evaluated at the end points of each polygonal segment (consistent with its 
traditional implementation).

It has been previously observed that averaged quantities (such as energy) 
converge at different rates, depending on the system, reference potential, 
and the form of the estimator \cite{FrDo:84,HeBrBe:82,ElDoCuFr:99}.  We use a 
virial estimator of the energy \cite{HeBrBe:82}, 
$E = \langle V(x) + x V'(x) / 2  \rangle$, which is known to exhibit improved 
convergence properties in many problems.  The accuracy of each average is
determined by comparing with the ``exact'' solution, computed by summing over 
the 15 lowest energy levels as calculated with Numerov's method \cite{Lev:00}.

\ifpreprintsty
\else
\begin{figure*}[thp]
\centering
\epsfysize=1.45in 
\epsfbox{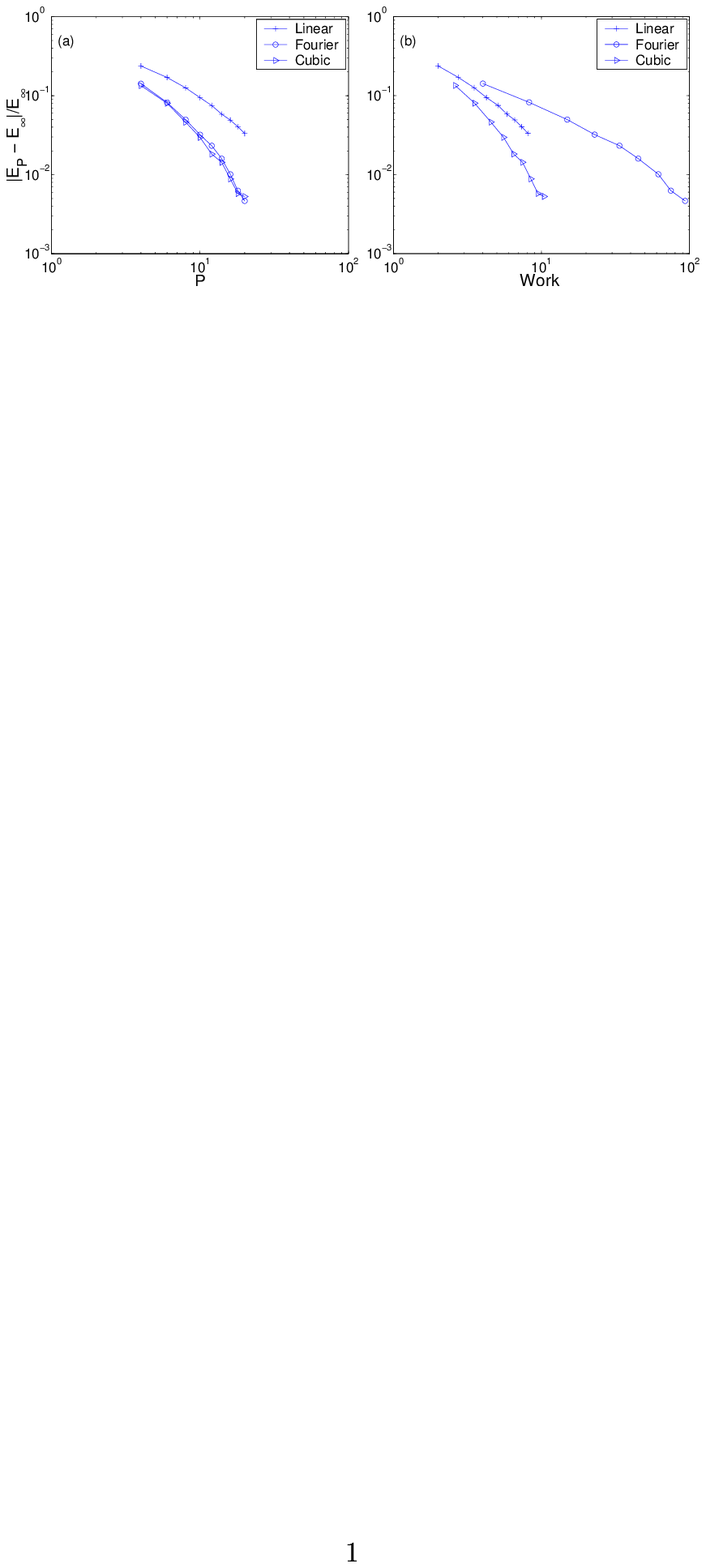}
\caption{Results for the double well. }
\label{experiments}
\end{figure*}
\fi

In Fig. \ref{experiments}, the error in the computed energy is shown as a
function of (a) the number of basis functions and (b) normalized CPU 
time.  When the number of basis functions (or potential evaluations) is used
as a measure of the work, we find that the FD and HCS methods are comparable,
and both are more efficient than the STA method.  However, when compared on 
the basis of CPU time, the HCS method is dramatically more efficient than 
both other methods.  The inefficiency of the FD method for low-dimensional
problems can be explained by considering the work required to compute $P$
points on the path.  This work scales like $O(P^2)$ for the FD method, since
the spectral basis functions are not compactly supported.   On the other hand,
for the STA and HCS methods this cost scales linearly with $P$.  Although
for very high dimensional problems, the cost of evaluating the potential should
dominate, and we expect that the differences in computational cost would not
be as pronounced.

In summary, the problem of approximating Feynman-Kac path integrals can be 
addressed using the finite-dimensional subspace approach.  This technique
allows for the formulation of new methods through the choice of a suitable
set of basis functions.  In addition, traditional methods such as the 
short-time approximation and Fourier discretization methods can be compared
using this framework.  As an example, by considering (Hermite) cubic splines, 
a new method can be constructed which exhibits improved efficiency when 
applied to a one-dimensional double-well problem.

The authors were supported by NSF Grant No. DMS-9627330.

\ifpreprintsty
\begin{figure*}[thp]
\centering
\epsfysize=3.0in 
\epsfbox{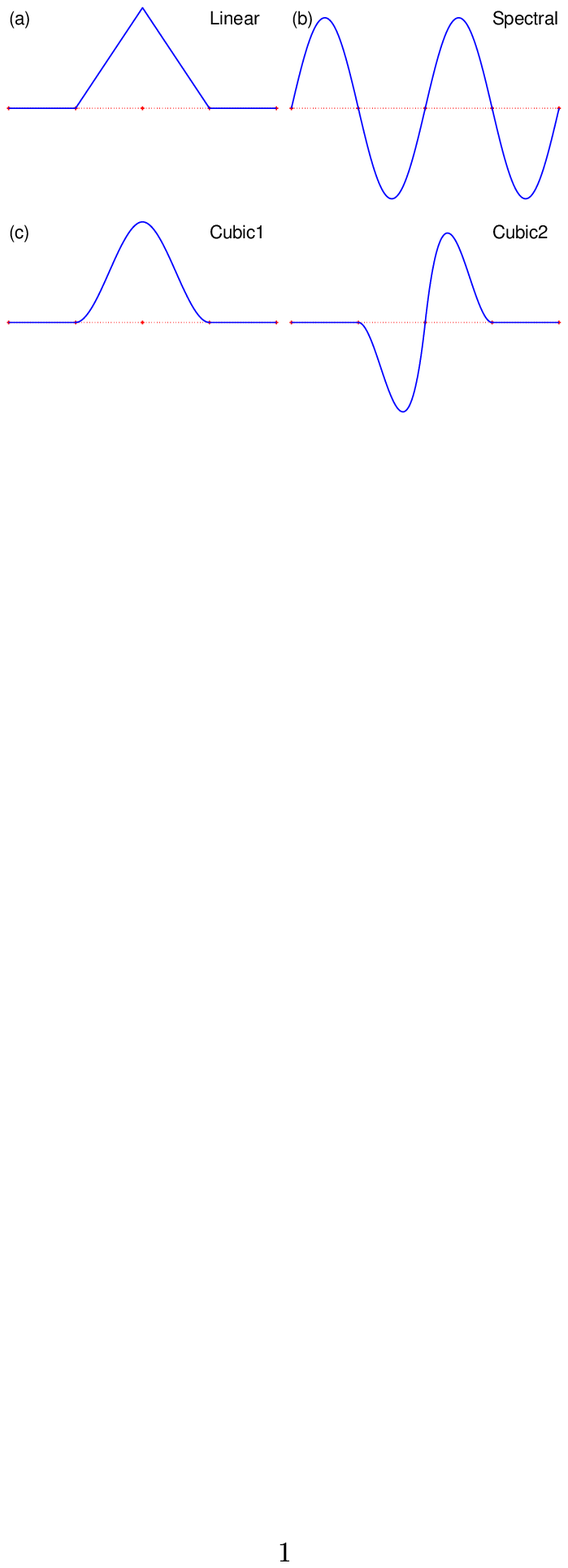}
\caption{Sample basis functions are shown above for the (a) linear, (b) spectral, and (c) cubic element methods.}
\label{elements}
\end{figure*}

\begin{figure*}[thp]
\centering
\epsfysize=2.5in 
\epsfbox{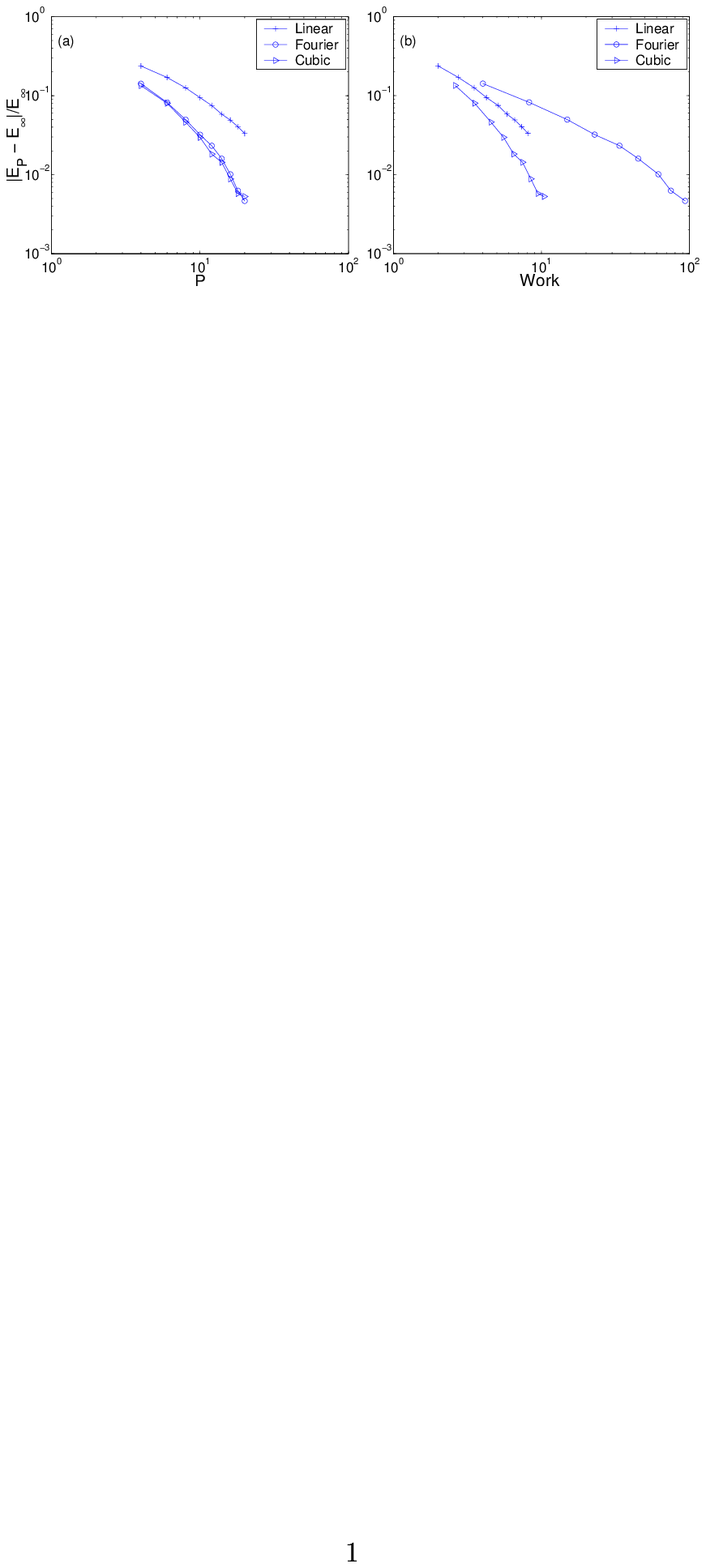}
\caption{Results for the double well. }
\label{experiments}
\end{figure*}
\fi


\begin{thebibliography}{10}

\bibitem{FeHi:65}
R.~P. Feynman and A.~R. Hibbs, {\em Quantum Mechanics and Path Integrals}
  (McGraw-Hill, New York, 1965).

\bibitem{Fe:72}
R.~P. Feynman, {\em Statistical Mechanics} (Benjamin, Reading, Mass., 1972).

\bibitem{ScStChWo:81}
K.~S. Schweizer, R.~M. Stratt, D. Chandler, and P.~G. Wolynes, J. Chem. Phys.
  {\bf 75},  1347  (1981).

\bibitem{Mi:75}
W.~H. Miller, J. Chem. Phys. {\bf 63},  1166  (1975).

\bibitem{FrDo:84}
D.~L. Freeman and J.~D. Doll, J. Chem. Phys. {\bf 80},  5709  (1984).

\bibitem{HeBrBe:82}
M.~F. Herman, E.~J. Bruskin, and B.~J. Berne, J. Chem. Phys. {\bf 76},  5150
  (1982).

\bibitem{DoCoFr:85}
J.~D. Doll, R.~D. Coalson, and D.~L. Freeman, Phys. Rev. Lett. {\bf 55},  1
  (1985).

\bibitem{CoFrDo:86}
R.~D. Coalson, D.~L. Freeman, and J.~D. Doll, J. Chem. Phys. {\bf 85},  4567
  (1986).

\bibitem{CoFrDo:89}
R.~D. Coalson, D.~L. Freeman, and J.~D. Doll, J. Chem. Phys. {\bf 91},  4242
  (1989).

\bibitem{DRDR:83}
H. {De Raedt} and B. {De Raedt}, Phys. Rev. A {\bf 28},  3575  (1983).

\bibitem{TaIm:84}
M. Takahashi and M. Imada, J. Phys. Soc. Jpn. {\bf 53},  3765  (1984).

\bibitem{MaMi:89}
N. Makri and W.~H. Miller, J. Chem. Phys. {\bf 90},  904  (1989).

\bibitem{MiSrTr:00}
S.~L. Mielke, J. Srinivasan, and D.~G. Truhlar, J. Chem. Phys. {\bf 112},  8758
   (2000).

\bibitem{Da:54}
B. Davison, Proc. Roy. Soc. London Ser. A {\bf 225},  252  (1954).

\bibitem{Ca:51}
R.~H. Cameron, Duke Math. J. {\bf 18},  111  (1951).

\bibitem{Co:86}
R.~D. Coalson, J. Chem. Phys. {\bf 85},  926  (1986).

\bibitem{StFi:73}
G. Strang and G.~J. Fix, {\em An Analysis of the Finite Element Method}
  (Prentice-Hall, Englewood Cliffs, N.J., 1973).

\bibitem{ElDoCuFr:99}
M. Eleftheriou, J.~D. Doll, E. Curotto, and D.~L. Freeman, J. Chem. Phys. {\bf
  110},  6657  (1999).

\bibitem{Lev:00}
I.~N. Levine, {\em Quantum Chemistry}, fifth ed. (Prentice Hall, Upper Saddle
  River, N.J., 2000).

\end{thebibliography}
\end{document}